\documentclass[runningheads]{llncs}
\usepackage{graphicx}
\usepackage{subcaption}
\usepackage{amssymb}
\usepackage{amsmath}
\usepackage{float}
\usepackage{verbatim}
\usepackage{glossaries}
\usepackage{booktabs}
\usepackage{multirow}
\usepackage{lscape}
\usepackage{arydshln}
\usepackage{array}
\usepackage[colorlinks=true,citecolor=blue,urlcolor=black]{hyperref}

\begin{document}

\title{Vertebrae Detection and Localization in CT with Two-Stage CNNs and Dense Annotations}
\titlerunning{Vertebrae Detection and Localization in CT with Two-Stage CNNs}
%
\author{James McCouat \and Ben Glocker}
\authorrunning{J. McCouat \and B. Glocker}
%
\institute{Department of Computing, Imperial College London, UK}
\maketitle              
\begin{abstract}
We propose a new, two-stage approach to the vertebrae centroid detection and localization problem. The first stage detects where the vertebrae appear in the scan using 3D samples, the second identifies the specific vertebrae within that region-of-interest using 2D slices. Our solution utilizes new techniques to improve the accuracy of the algorithm such as a revised approach to dense labelling from sparse centroid annotations and usage of large anisotropic kernels in the base level of a U-net architecture to maximize the receptive field. Our method improves the state-of-the-art's mean localization accuracy by 0.87mm on a publicly available spine CT benchmark.
\end{abstract}

\section{Introduction}
Medical imaging to capture the spine is a key tool used in clinical practice for diagnosis and treatment of patients suffering from spinal conditions. A common image modality used in clinics is Computed Tomography (CT), which provides a detailed view of the spinal anatomy. Once a scan of a patient is acquired the scan must be analyzed, a task which is usually performed by a radiologist. One such analysis would be to locate and identify which vertebrae are visible within the scan. However, this is not a trivial task, is error-prone and can be time-consuming. First, as CT scans use X-rays the exposure to the patient is often limited by only scanning the part of the body which is of interest. This means that a limited number of vertebrae are captured in each scan. These scans, known as arbitrary Field-Of-View (FoV) scans, make the vertebrae harder to identify because a radiologist cannot simply count from the end of the column or because they miss important visual context. Second, the scans often include severe pathological cases which could mean that the spine is an abnormal shape or the scan could be post-op and contain metal implants causing imaging artifacts. Third, many vertebrae have similar appearance and are hard to tell apart without contextual, anatomical information. \par
An algorithm which could perform the task of vertebrae localization and identification could not only inform radiologists and speed up their workflow, it could also be used as prior information for subsequent algorithms, for example, a lesion or fracture detection algorithm. In addition, automation could be used to perform a survey over a large dataset of spines to record information about a population, a task which is difficult to do manually by human experts. \par
The same reasons that make this problem hard for radiologists also make it difficult for an algorithm. In 2012, Glocker \textit{et al}. \cite{glocker2012automatic}. proposed a solution to the problem which regressed the centroid positions however the method made assumptions about the shape of the spine and thus did not work well on pathological cases. Later, in \cite{glocker2013vertebrae} this was improved upon from their previous solution by turning the problem from a regression into a dense classification problem. This was done by generating a dense labelling, which is a label given to each pixel in the scan, from the ground-truth centroid annotations. The dense labelling is then converted back to the sparse labelling of centroid positions at the end of the processing pipeline. This approach worked much better for pathological cases. Both approaches \cite{glocker2012automatic,glocker2013vertebrae} used random forests but there has since been an emergence of convolutional neural networks (CNN) in medical image analysis \cite{dou2016automatic,liu2018automatic,roth2018application} and better results have been achieved for the vertebrae localization task with deep learning \cite{chen2015automatic}. Some use a U-net architecture, such as Yang \textit{et al} \cite{yang2017automatic},  which is a popular architecture for image segmentation problems \cite{ronneberger2015u}. In 2018, Liao \textit{et al}. \cite{liao2018joint} published a state-of-the-art solution which regresses the positions of the centroids using a CNN combined with a Recurrent Neural Network (RNN) to capture the ordering of the vertebrae and to incorporate long-range contextual information. Based on the literature, we decided to develop a new approach which should have the following features: 1) Use CNNs, 2) Capture the ordering of the vertebrae to improve accuracy, 3) Use a Dense Labelling strategy and turn the problem into a classification task, 4) Capture short and long-range contextual information from the scan.

\section{Method}
\begin{figure}
    \begin{subfigure}[c]{0.24\textwidth}
      \centering
        \includegraphics[width=1.0\linewidth]{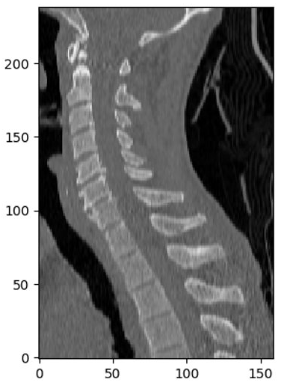}
        \caption{}
        \label{fig:initial}
    \end{subfigure}
    \begin{subfigure}[c]{0.24\textwidth}
      \centering
        \includegraphics[width=1.0\linewidth]{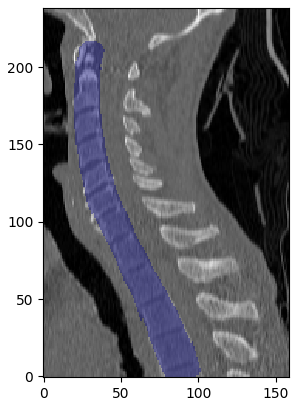}
        \caption{}
        \label{fig:detection}
    \end{subfigure}
    \begin{subfigure}[c]{0.24\textwidth}
      \centering
        \includegraphics[width=1.0\linewidth]{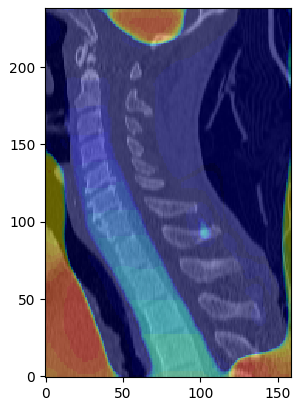}
        \caption{}
        \label{fig:straight_identification}
    \end{subfigure}
    \begin{subfigure}[c]{0.24\textwidth}
      \centering
        \includegraphics[width=1.0\linewidth]{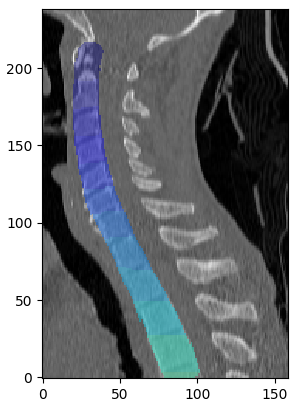}
        \caption{}
        \label{fig:final_identification}
    \end{subfigure}
    \caption{(a) shows an original scan in greyscale, (b) shows the output of the detection model applied to the scan, (c) shows the output of the identification model applied to the scan, (d) shows (b) and (c) multiplied together to produce a final prediction for each pixel.}
\end{figure}

Our method uses two CNNs, first a detection model which segments the vertebrae from the background (see Fig. \ref{fig:detection}), then an identification model which identifies which pixels belong to which vertebra (see Fig. \ref{fig:straight_identification}). This approach was arrived at by analyzing how we could capture the ordering of the vertebrae. The identification model does not classify each pixel discretely but instead produces a continuous value for each pixel. This value is then rounded to an integer which corresponds to a specific vertebra, for example 4 = C4 vertebra. This means we can use an L1 loss function on each pixel and thus capture the ordering through this loss function. This approach would not have been possible if we had included background pixels in the identification stage therefore we segment out the background in the detection step. Our approach also captures short-range and long-range information, 3D short-range information is captured in the detection model by feeding in small 3D samples to train the network. The identification model is trained by feeding in large slices which capture long-range information, which is essential for the task of identifying individual vertebrae. To produce our final dense predictions we multiply the results of the detection model and identification model to produce a labelling on each pixel (see Fig. \ref{fig:final_identification}). These final predictions are then aggregated to produce final centroid estimates for each vertebra. 

\subsection{Detection Model}

\begin{figure}
    \centering
    \includegraphics[width=1.0\textwidth]{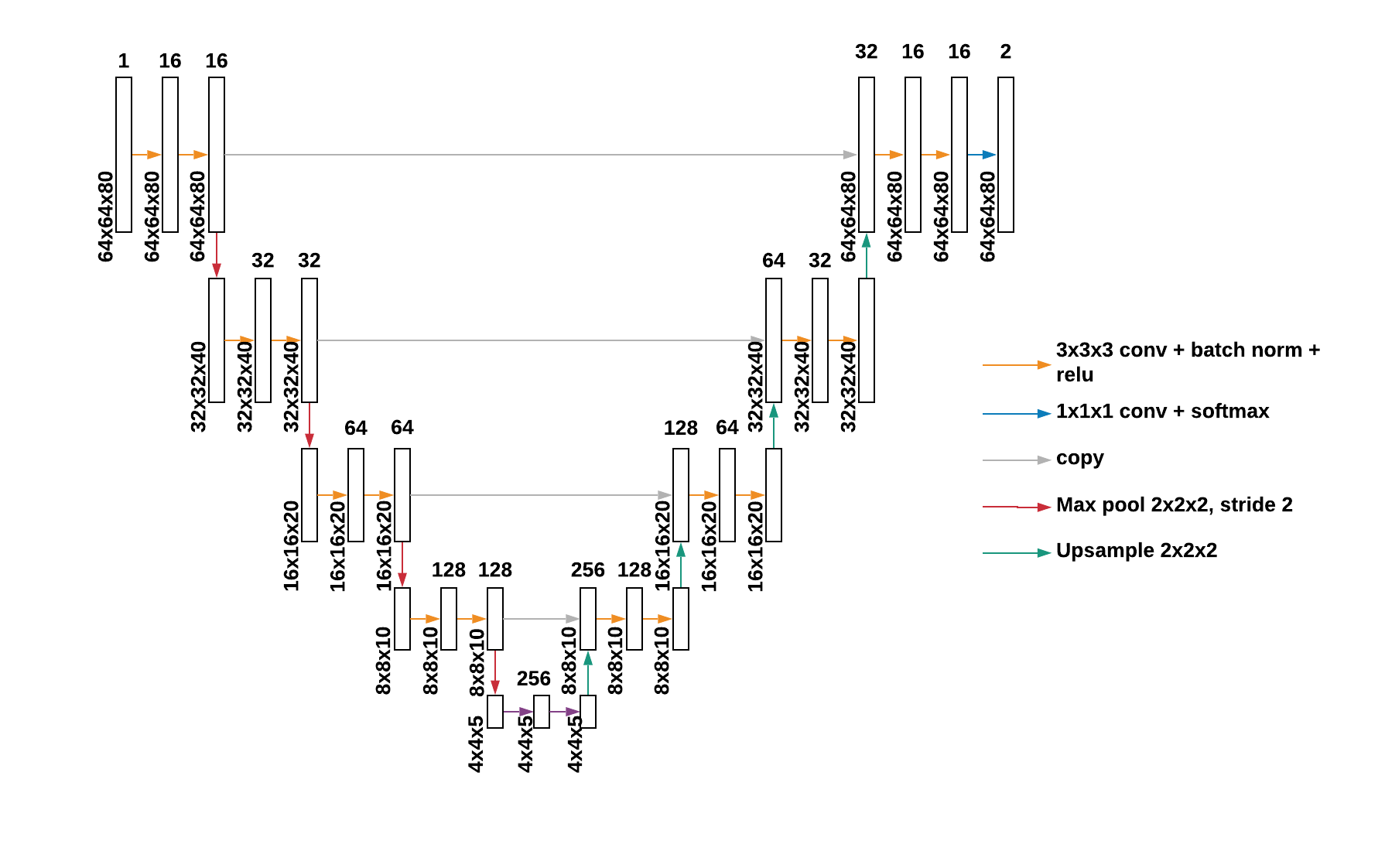}
    \caption{Detection Model Architecture}
    \label{fig:detection_architecture}
\end{figure}

The detection model classifies each pixel of a scan as either `background' or `vertebrae'\footnote{Specifically 'vertebrae' means 'vertebral bodies' because the posterior components of the vertebrae are not segmented out.}. To train the network we feed in 5 cropped samples for each scan in the training set (enforcing that at least 4 of them contain some vertebrae pixels), each sample has size 64 x 64 x 80 pixels. Each sample has an accompanying dense labelling, containing 0s (for background) and 1s (for vertebrae), of the same size, for the network to learn from. A 3D U-net architecture was used as portrayed in Fig. \ref{fig:detection_architecture}. We used a weighted categorical cross entropy loss with 2 classes as our loss function. We gave the background label a 0.1 weighting and the vertebrae label a 0.9 weighting to reflect the proportion of background labels in the samples. The full loss function is shown in Eq. (\ref{eq:detec_loss}). 

\begin{equation} \label{eq:detec_loss}
L(P, Q) = 0.1 * P(0)\log Q(0) + 0.9 * P(1)\log Q(1)
\end{equation}
\\
\noindent
We used 'same' padding for all convolutional layers with stride 1, a learning rate $\lambda$ of 0.001, a batch size of 16, batch normalization after every convolutional layer with momentum of 0.1 and trained for 50 epochs which took 11 hours on our hardware (see Section. \ref{sec:experiments}). We obtained a validation Dice score of 0.961 on the 1 (vertebrae) labels and a validation accuracy of 98.5\% on test samples generated from our test set. \par 
At test time we applied the detection model to a scan patch-wise with some overlap between patches. The input to the network is 64 x 64 x 80 and we applied it in steps of 32 x 32 x 40, padding the border of the scan by 16 x 16 x 20 and discarding the outer border of size 16 x 16 x 20 from each output. This reduced edge artifacts in the detection and led to improved mean localization scores.

\subsection{Identification Model}
The identification model outputs a continuous value for each pixel of a scan which, is rounded to an integer to correspond to a specific vertebrae. The identification model gives a value to each pixel even if that pixel is not depicting any vertebrae and is a background. These pixels will be filtered out of the final predictions by multiplying them by their corresponding pixel from the output of the detection model which should be 0 for a background pixel. To train the identification network we feed in cropped samples of size 8 x 80 x 320. We generate 100 cropped samples for each scan in the training set (enforcing that all of them capture some vertebrae pixels), each with a corresponding dense labelling of size 80 x 320, representing the labelling for the 4th slice of the input sample. We also elastically deform each of these samples along the 2nd and 3rd axis using the elastic-deform python package\footnote{\url{https://pypi.org/project/elasticdeform/}} (with $\sigma = 0.7$ on a $3 \times 3$ grid). \par
\begin{figure}
    \centering
    \includegraphics[width=1.0\textwidth]{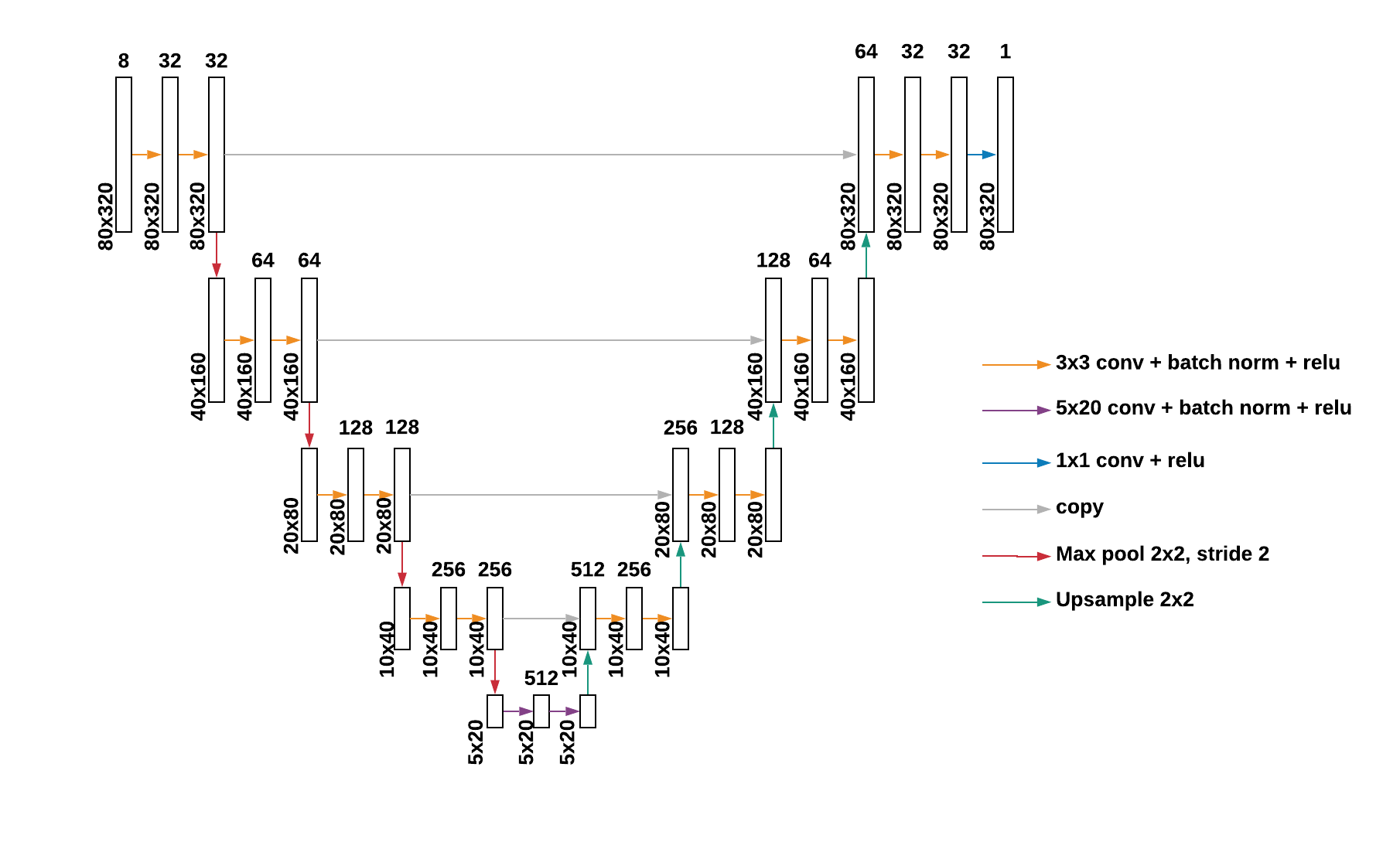}
    \caption{Identification Model Architecture}
    \label{fig:identification_architecture}
\end{figure}

We used a 2D U-net architecture for the identification model with 8 channels in the input layer to pass in our samples of size 8 x 80 x 320. We implemented a U-net architecture with large anisotropic filters of size 5 x 20 at the lowest level of the architecture to increase the size of the receptive field thus maximizing the contextual information captured by the network (see Fig. \ref{fig:identification_architecture}). As mentioned previously, pixels which are background will be filtered out by multiplying by the detection model. This means that we do not care about the pixels which are labelled as background when we train this model. This is reflected in the loss function which is a modified L1 loss function given below:

\begin{equation}
    \label{id_loss_function}
    L=
    \begin{cases}
      |y_i - x_i| & \text{if}\ x_i \neq 0 \\
      0, & \text{otherwise}
    \end{cases}
 \end{equation}
 where $y_i$ is the predicted value of pixel i and $x_i$ is the true value of pixel i.\\
 \\
 We used 'same' padding for all convolutional layers with stride 1, a learning rate $\lambda$ of 0.001, a batch size of 32, batch normalization after every convolutional layer with momentum of 0.1 and trained for 35 epochs which took 7 hours. \par
 The identification model is a fully convolutional network \cite{long2015fully} which allows us to apply it to whole slices of the input scan, padded to the nearest multiple of 16, at test time.

\subsection{Sparse to Dense Annotation} \label{sec:sparse_to_dense}
\begin{figure}
    \begin{subfigure}[c]{0.4\textwidth}
      \centering
        \includegraphics[width=0.75\linewidth]{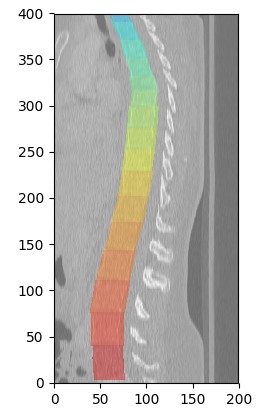}
        \caption{}
        \label{fig:complete_dense_labelling}
    \end{subfigure}
    \begin{subfigure}[c]{0.3\textwidth}
      \centering
        \includegraphics[width=0.9\linewidth]{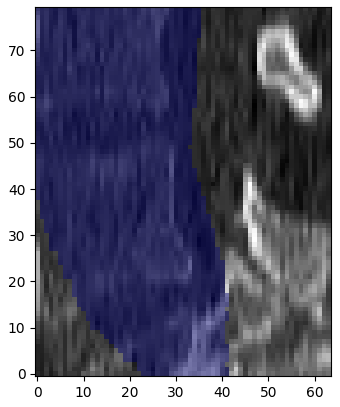}
        \caption{}
        \label{fig:detection_sample}
    \end{subfigure}
    \begin{subfigure}[c]{0.25\textwidth}
      \centering
        \includegraphics[width=0.75\linewidth]{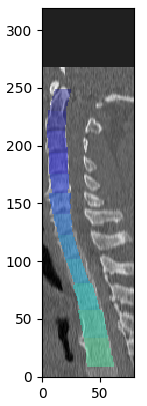}
        \caption{}
        \label{fig:identification_sample}
    \end{subfigure}
    \caption{(a) shows a dense labelling, (b) shows an example of a sample used to train the detection model (the sample is 64 x 64 x 80 but only a 64 x 80 slice is shown here). Note: any value over 1 in the dense labelling is converted to 1 for the detection model samples, (c) shows an example of a sample used to train the identification model. Note: The size of the sample is 8 x 80 x 320, if the original scan is not larger enough to fill those dimensions some padding is added, as can be seen at the top.}
\end{figure}

To train both the detection and the identification model we feed in samples each with a corresponding dense labelling. In the case of the detection model the dense labelling contains two values; 0 representing background and 1 representing vertebrae (see Fig. \ref{fig:detection_sample}). The identification model's dense labelling contains values from 0 to 26, 0 representing background, 1 representing C1 vertebrae etc... up to 26 representing S2 vertebrae (see Fig. \ref{fig:identification_sample}). Our dataset comes with centroid positions (sparse labels) which must be converted to dense labels. In previous papers a solution was to plot spheres around each of the centroids \cite{glocker2013vertebrae}. This approach however is not the most accurate dense labelling because vertebrae are not spherical. We implemented an algorithm which produces an anatomically better dense labelling, which is to say a better approximation of which pixels are vertebrae (detection) and which vertebra those pixels correspond to (identification), from the ground-truth sparse annotations. Our algorithm is as follows:

\begin{enumerate}
    \item Find midpoints between all adjacent centroids in the column.
    \item Draw line segments between these midpoints. Add additional line segments at the start and end of the column so that there is a line segment to represent each vertebra. 
    \item Plot discs (on the plane of the sagittal and transverse axes) around each point on the line segment. The radius of these discs is specific to the vertebra the line segment represents. We obtained approximations of the radii for specific vertebrae from Busscher \textit{et al}. \cite{busscher2010comparative}.
\end{enumerate}

\noindent
A scan annotated with a dense labelling produced by this method is shown in Fig. \ref{fig:complete_dense_labelling}.

\subsection{Dense to Sparse Annotation}
Once a label has been predicted for each pixel (see Fig. \ref{fig:final_identification}) these labels are aggregated to calculate predicted centroid positions. To aggregate we find the median position of all pixels which vote for each vertebra. However before calculating this we apply a threshold so that if there are less than a certain number of pixels voting for a vertebra we do not include that centroid in the prediction. This filters out erroneous predictions produced by the identification model. The threshold $x_{v}$ is specific to the vertebra $v$ and is calculated using Eq. \ref{eq:theshold}.

\begin{equation} \label{eq:theshold}
    x_{v}=max(3000, 0.4R_{v}^{3})
 \end{equation}
 where R is the radius of the vertebra v (see Section. \ref{sec:sparse_to_dense}).\\

\section{Experiments} \label{sec:experiments}
The dataset we used for experiments is a public dataset provided by the BioMedia research group\footnote{\url{https://biomedia.doc.ic.ac.uk/data/spine/}}. This dataset is split into a dedicated training set of 242 scans and a dedicated test set of 60 scans. This is the same dataset which has been used in previous work allowing us to directly compare results (see Table \ref{tab:main_results}). It is comprised of pathological, arbitrary FoV scans. We re-sample the scans at 1mm x 1mm x 1mm resolution so every pixel in the samples we generate from the scans represents a 1mm cubed section of the body. We used a Microsoft Azure Virtual Machine with a NVIDIA Tesla K80 and 56GB of main memory to conduct our experiments. Source code for this project is available on GitHub which can be used to reproduce the results we achieved\footnote{\url{https://github.com/jfm15/SpineFinder}}.

\subsection{Results}
The results reported in this section are calculated over the dedicated testing set of 60 scans. We measured metrics, Id rate, mean localization distance and standard deviation (std) distance. Id rate is the percentage of the centroid estimations predicted which are closest to the correct ground truth vertebra centroid (and are less than 20mm from that centroid). Mean and Std relate to the localization error distance between the predicted centroid positions and the ground-truth centroid positions for the same vertebrae (if it occurs in the scan). Table \ref{tab:main_results} shows a comparison between our method and 2 of the latest methods (Liao \textit{et al}. \cite{liao2018joint} being the state-of-the-art).

\begin{table}
\begin{center}
\caption{Comparison of results}\label{tab:main_results}
\begin{tabular}{ c@{\hskip 12pt}ccc@{\hskip 12pt}ccc@{\hskip 12pt}ccc } 
 \cline{1-10}
 \multicolumn{1}{c}{} &
 \multicolumn{3}{c}{Chen et al. \cite{chen2015automatic}} &
 \multicolumn{3}{c}{Liao et al. \cite{liao2018joint}} &
 \multicolumn{3}{c}{Our Method} \\
 \cline{1-10}
 Region & Id Rate & Mean & Std & Id Rate & Mean & Std & Id Rate & Mean & Std \\

 All & 84.2\% & 8.82 & 13.04 & \textbf{88.3}\% & 6.47 & 8.56 & 85.8\% & \textbf{5.60} & \textbf{7.10} \\ 

 Cervical & 91.8\% & 5.12 & 8.22 & \textbf{95.1}\% & 4.48 & \textbf{4.56} & 90.6\% & \textbf{3.93} & 5.27 \\ 

 Thoracic & 76.4\% & 11.39 & 16.48 & \textbf{84.0}\% & 7.78 & 10.17 & 79.8\% & \textbf{6.61} & \textbf{7.40} \\ 

 Lumbar & 88.1\% & 8.42 & 8.62 & \textbf{92.2}\% & 5.61 & \textbf{7.68} & 92.0\% & \textbf{5.39} & 8.70 \\ 
 \cline{1-10}
\end{tabular}
\end{center}
\end{table}

Table \ref{tab:main_results} shows that our method improves on the mean localization error produced by the state-of-the-art (Liao \textit{et al}. \cite{liao2018joint}) by 0.87mm, a good margin. It performs slightly worse on the Id rate. A possible reason for this is that our method relies on the ordering of vertebrae and sufficient contextual information to localize in a dense regression identification stage, but it does not explicitly classify vertebrae, for example, using categorical cross entropy as in Liao \textit{et al}. \cite{liao2018joint}. \par
As well as improved mean localization, an advantage of our method is that it is lightweight and fast both in training time, models took 11 hours (detection model) and 7 hours (identification model) to train, but also at test time, with predictions for a scan being computed on average in 40 seconds. Our method does not rely on an iterative stage, unlike Liao \textit{et al}. \cite{liao2018joint} which uses an RNN, thus making our method likely more efficient at test time.

\begin{figure}
    \centering
    \includegraphics[width=1.0\textwidth]{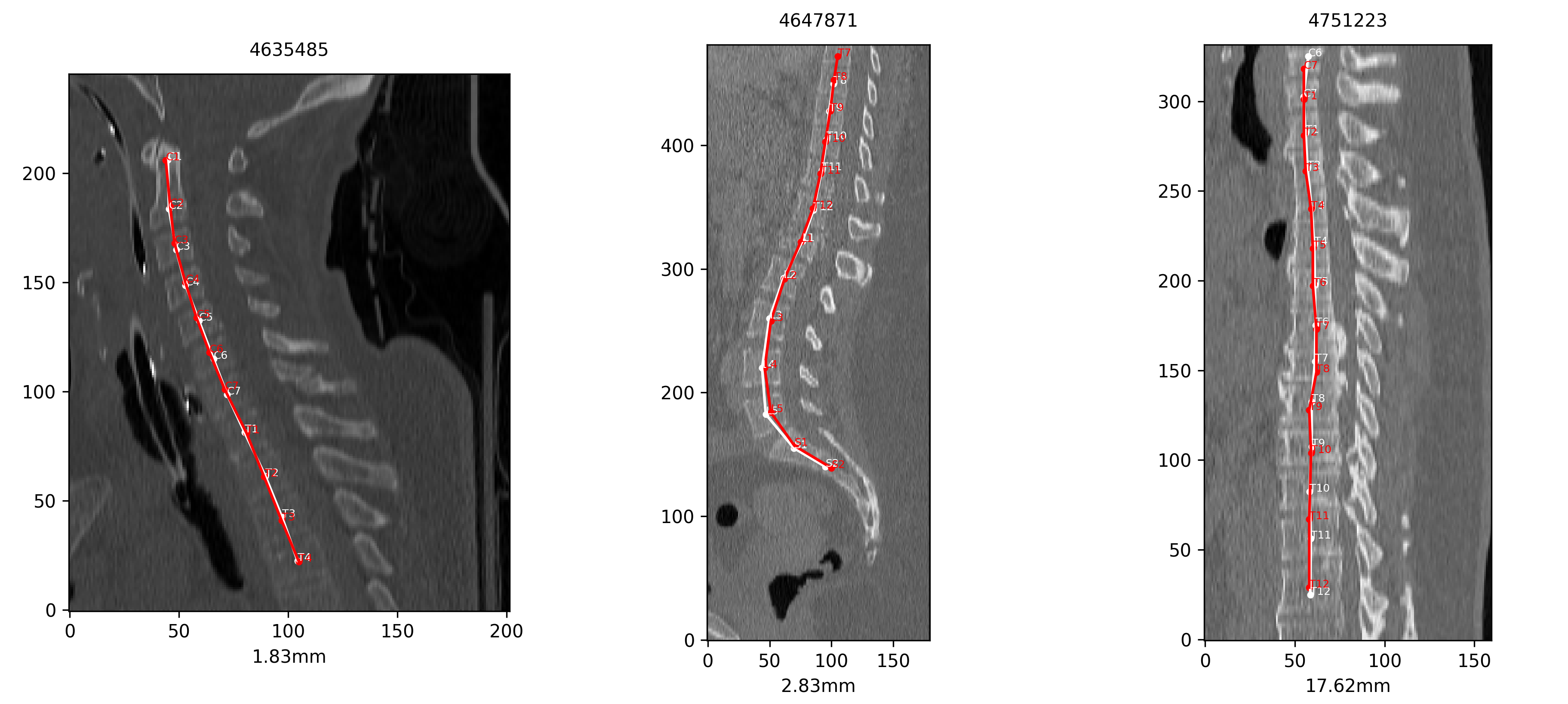}
    \caption{Example of 3 predictions, the red points are our method's centroid predictions, the white points are the ground-truth centroid locations, below each scan the mean localization error for that scan is shown.}
    \label{fig:cases}
\end{figure}

\subsection{Analysis of Worst Cases}

\begin{figure}
    \centering
    \includegraphics[width=1.0\textwidth]{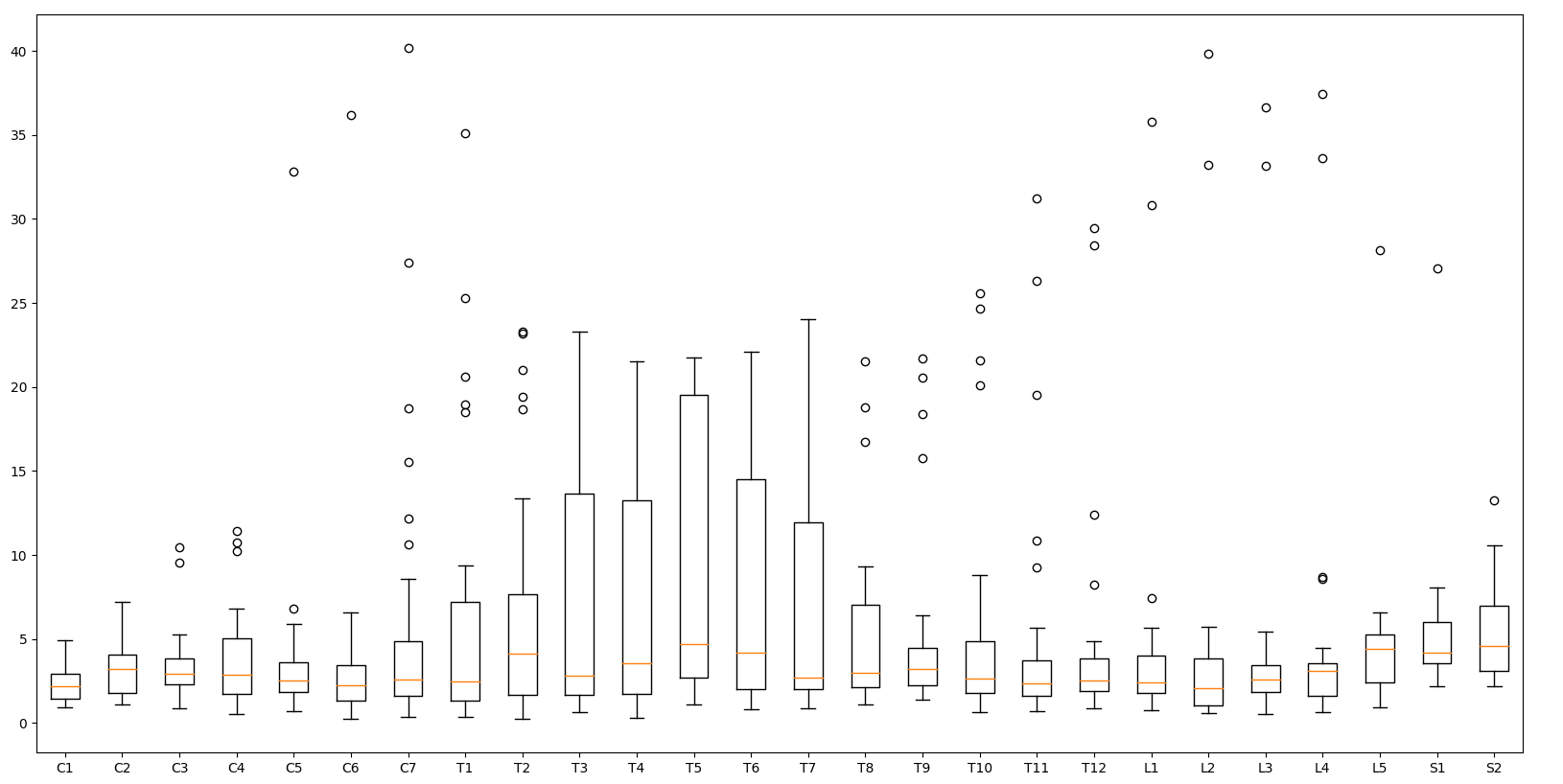}
    \caption{Mean localization error per vertebra}
    \label{fig:box_plot}
\end{figure}

In Fig. \ref{fig:cases} the left-most two examples show very accurate predictions, the right-most shows an example of one of the least accurate predictions in the test set reporting an average error of 17.62mm. This scan shows a weakness in our method where in scans which mainly capture the thoracic vertebrae, our method can predict the entire column inaccurately. It is common in vertebrae localization to be least effective on the thoracic vertebrae as shown in Table \ref{tab:main_results} and the plot in Fig \ref{fig:box_plot}. A possible reason is that the ends of the vertebrae column are rarely captured in scans with lots of thoracic vertebrae thus the algorithm cannot simply count up or down the column. In addition the inaccuracy with thoracic vertebrae for our method could be due to them being under-represented in the dataset. T6 and T7 have the least representation with 80 and 82 scans containing them, out of 242. Unlike other methods we do not sample the vertebrae evenly, instead we take cropped samples from the scans at random positions. This is something which could be considered to improve our results in future work.

\section{Conclusion}
Using a two stage CNN approach, first to detect vertebrae with a 2-class cross entropy loss, then a second stage to identify the specific vertebrae capturing the ordering using an L1 loss, has been shown to provide an effective solution to the vertebrae localization problem and improves upon the state-of-the-art's mean localization accuracy by a good margin on a publicly available pathological spine CT dataset. In future these results could possibly be improved by evenly sampling the vertebrae and by feeding even more contextual information into the identification model or by enforcing the use of surrounding tissue to improve thoracic vertebrae results. Other directions could be to optimize the employed network architectures and explore robust centroid estimate techniques such as mean shift. Another interesting direction could be to incorporate more explicitly shape and ordering constraints of the spinal anatomy.

\bibliography{paper}
\bibliographystyle{splncs04}

\end{document}